\documentclass[12pt,preprint]{aastex}

\begin{document}


\title{Measuring Magnetic Fields in Ultracool Stars and Brown Dwarfs}


\author{A. Reiners\altaffilmark{1,^\star} \and G. Basri}
\affil{Astronomy Department, University of California, Berkeley, CA 94720
\email{[areiners, basri]@astron.berkeley.edu}}
\altaffiltext{1}{Hamburger Sternwarte, Universit\"at Hamburg, Gojenbergsweg 112, D-21029 Hamburg, Germany}
\altaffiltext{$^\star$}{Marie Curie Outgoing International Fellow}


\begin{abstract}
  We present a new method for direct measurement of magnetic fields on
  ultracool stars and brown dwarfs. It takes advantage of the
  Wing-Ford band of FeH, which are seen throughout the M and L
  spectral types. These molecular features are not as blended as other
  optical molecular bands, are reasonably strong through most of the
  spectral range, and exhibit a response to magnetic fields which is
  easier to detect than other magnetic diagnostics, including the
  usual optical and near-infrared atomic spectral lines that have
  heretofore been employed. The FeH bands show a systematic growth as
  the star gets cooler. We do not find any contamination by CrH in the
  relevant spectral region. We are able to model cool and
  rapidly-rotating spectra from warmer, slowly-rotating spectra
  utilizing an interpolation scheme based on optical depth scaling.
  We show that the FeH features can distinguish between negligible,
  moderate, and high magnetic fluxes on low-mass dwarfs, with a
  current accuracy of about one kilogauss. Two different approaches to
  extracting the information from the spectra are developed and
  compared.  Which one is superior depends on a number of factors. We
  demostrate the validity of our new procedures by comparing the
  spectra of three M stars whose magnetic fluxes are already known
  from atomic line analysis. The low and high field stars are used to
  produce interpolated moderate-strength spectra which closely
  resemble the moderate-field star. The assumption of linear behavior
  for the magnetic effects appears to be reasonable, but until the
  molecular constants are better understood the method is subject to
  that assumption, and rather approximate. Nonetheless, it opens a new
  regime of very low-mass objects to direct confirmation and testing
  of their magnetic dynamos.
\end{abstract}
\keywords{stars: low mass, brown dwarfs --- line: profiles --- stars: magnetic fields}

\newpage

\section{Introduction}

It is generally accepted that magnetic fields are responsible for the
generation of solar and stellar activity. In the Sun, magnetic fields
can be directly associated with active regions in spatially resolved
images. From observations of solar-like stars, the rotation-activity
connection was established; the rate of activity grows with faster
rotation, which is explained by the rotational sensitivity of a 
dynamo. In solar-type stars, this dynamo is presumed to
lie at the interface between the star's radiative core and the
convective envelope. No break in activity has been found at the
transition to fully convective stars, but a different type of dynamo
must be working in such cases. The mean level of activity does show 
a rapid falloff around spectral type M9, even in rapid
rotators \citep{Basri95, Mohanty03}. Some of these objects have 
extremely low Rossby numbers, for which one could imagine that the
dynamo could be inhibited. Magnetic fields in very cool atmospheres 
are also expected to suffer from very large electrical resistivities 
and thus efficient field diffusion \citep{Mohanty02}. This can prevent
atmospheric motions from being converted into non-radiative heating
through magnetic and current dissipation (or MHD wave generation);
these mechanisms are thought to be the source of much of the 
stellar activity in warmer stars.

It has been found that ultracool stars and brown dwarfs of spectral 
types late M and L can sometimes support low levels of quiescent stellar 
activity \cite[e.g.][]{Basri00, Gizis02, West04}. It is also the 
case that stellar flares are observed in some of the ultracool objects
\cite[e.g.][]{Reid99, Schmitt02, Liebert03, Stelzer04}, although they 
seem to be different than in the solar case, at least in their ratio of 
radio to X-ray luminosities \cite[e.g.][]{Rutledge00, Berger02, Berger05}. 
In objects M9 and cooler, it is still an open question how much 
of the observed photometric variability is due to clouds vs. starspots, 
but H$\alpha$\ emission and flaring must be a result of magnetic fields.
How is magnetic activity and flaring generated in these objects? 

Stellar magnetic fields are usually measured through magnetic Zeeman
broadening in atomic lines that have large Land\'e\,$g$ values
\citep[e.g.][and references therein]{Robinson80, Marcy89, Saar01,
  Solanki91, Saar96}. The measurement is usually carried out by
comparing the profiles of magnetically sensitive and insensitive
absorption lines between observations and model spectra. An alternate
method that relies on the change in line equivalent widths has also
been developed by \cite{BMV92}. Modeling in both cases requires the
use of a polarized radiative transfer code and knowledge of the Zeeman
shift for each Zeeman component in the magnetic field.  Furthermore,
it requires the observed lines to be isolated and that they can be
measured against a well-defined continuum. The latter becomes more and
more difficult in cooler stars since atomic lines vanish in the
low-excitation atmospheres and among the ubiquitous molecular lines
that appear in the spectra of cool stars.  Measurements of stellar
magnetic fields extend to stars as late as M4.5 \citep{JKV96, JKV00,
  Saar01}. In cooler objects, atomic lines can no longer be used for
the above-mentioned reasons, and because suitable lines become
increasingly rare.

One idea to overcome this problem is to look for Zeeman broadening in
molecular lines in ultracool objects. In the sunspot atlas of
\cite{Wallace98}, the strong magnetic sensitivity of the Wing-Ford
band of FeH just before 1\,$\mu$m is clearly demonstrated.
\cite{Valenti01} suggested that FeH would be a useful molecular
diagnostic for measuring magnetic fields on ultracool dwarfs, but they
point out that improved laboratory or theoretical line data are
required in order to model the spectra directly.

In this paper we investigate the possibility of detecting (and
measuring) magnetic fields in FeH lines of ultracool dwarfs through
comparison between the spectrum of a star with unknown magnetic field
strength and spectra of stars for which the magnetic field strength can
be calibrated in atomic lines. Such calibrators are generally early-
to mid- type M-dwarfs. We explain what sort of data are required at the
beginning of \S\ref{sect:FeH}, then go on to investigate the behavior
of the FeH absorption band in stars of different spectral types. A
closer look at the magnetic sensitivity of individual FeH lines is
accomplished by comparing the spectrum of an active star with a strong
magnetic field that was measured in atomic lines to a spectrum of an
inactive star that shows no signatures of a magnetic field
(\S\ref{sect:FeH_magnetic}). Line ratios of magnetically sensitive and
insensitive lines are set forth in \S\ref{sect:detectability}; they are
used to determine the detectability of magnetic signatures in stars of
different spectral type and rotation velocity. As an example, we
determine the magnetic field in a star for which the field is already
known from an atomic line analysis using two approaches, FeH line
ratios (\S\ref{sect:Bfratio}) and direct $\chi^2$-fitting to the FeH spectrum 
(\S\ref{sect:Bfchisq}). We give a summary in \S\ref{sect:Summary}.

\section{The Wing-Ford band of FeH in Ultracool Stars}
\label{sect:FeH}

The data we use here were taken during three nights in March and
August 2005 with the HIRES spectrograph at Keck~I \citep{Vogt94}.  The
observations were carried out after the detector upgrade, which
replaced the previous chip with three new ones. This extends the
accessible spectral range and significantly improves the sensitivity
in the near infrared. We were able to obtain spectra that cover the
wavelength region from H$\alpha$ up to 1\,$\mu$m with one exposure,
although some gaps occur between the order in the red. Since we
observe relatively faint objects along with bright ones, we chose a
slit width of 1.15\,arcsec, which yields a resolution of $R =
31\,000$. We make use of the FeH bands in mid-type M stars, which are
relatively bright. The signal-to-noise ratio (SNR) we achieved in
those is up to 100 around 1\,$\mu$m. The data were reduced using the
MIDAS reduction software, they were flat-fielded and filtered for
cosmic rays. We subtracted light from sky emission lines by
individually extracting the sky spectrum.

Strong absorption around 9900\,\AA\ in spectra of M-dwarfs was first
observed by \citet{WF69}. \cite{Nordh77} and \cite{Wing77}
compared spectra of a sunspot and of M-dwarfs to laboratory spectra
and assigned the absorption to FeH, and \cite{Balfour83} established
the Wing-Ford band as a ${^4\Delta} - {^4\Delta}$ transition of FeH.
The rotational analysis of the ${^4\Delta} - {^4\Delta}$ system was
given by \cite{Phillips87}, and \cite{Dulick03} provided new line
lists and opacities for this system. The band around 9900\,\AA\ arises
from the $0-0$ $(v'-v'')$ transition of the
${F\,^4\Delta}-{X\,^4\Delta}$ system.  It is already visible in stars
of spectral type K5 \citep{Valenti01} and strengthens through spectral
class M to the early L-dwarfs before it fades through the late L
dwarfs and early T dwarfs \citep{McLean03}. Fading of the bandhead can
be explained by iron condensing out of the atmosphere. The bandhead
reappears in early to mid T dwarfs as shown by \citet{Burgasser02}. To
explain this strengthening, the authors propose holes to be present in
the cloud layer that allow one to see the deeper and hotter layers where
FeH is not depleted.

\cite{Schiavon97} studied the dependence of the Wing-Ford band on
model atmosphere parameters, comparing them to medium resolution ($R =
13\,000$) spectra of mid-type M stars. Their model spectra match the
observed spectra only very coarsely, which may in part be due to
their choice of oscillator strengths. Nevertheless, at this
resolution, their model seems to produce absorption features at all
positions where absorption is visible in the data, thus indicating
that FeH is the only major opacity source in that wavelength region.
Proceeding towards cooler stars, \cite{Cushing03} compared low
resolution ($R = 2000$) spectra of a late-type M and three L dwarfs to
a laboratory spectrum of FeH using a King furnace at $T = 2700$\,K.
They identify 34 FeH features that are visible in stellar and
laboratory spectra.

\begin{figure}
  \plotone{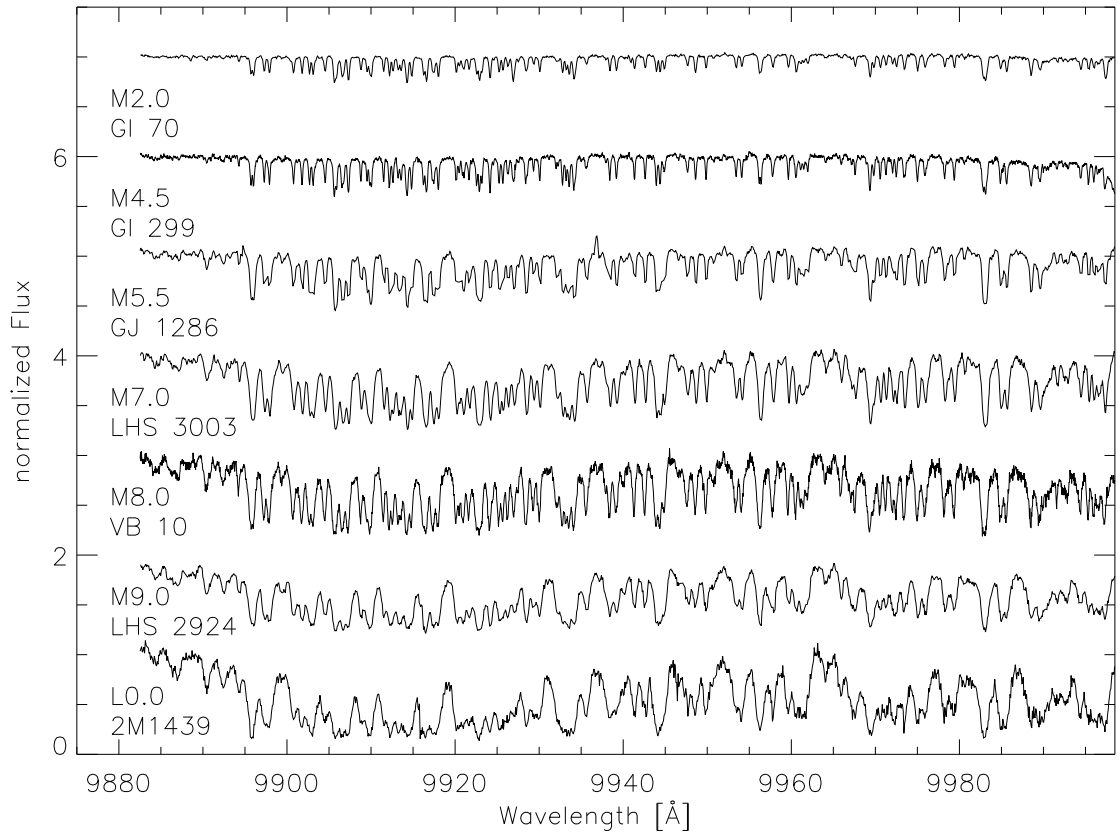}
  \caption{\label{fig:sequence}The Wing-Ford band head in objects of spectral types M2 to L0. }
\end{figure}

We show a high resolution spectroscopic sequence of the Wing-Ford band
in the spectral types M2\,--\,L0 in Fig.\,\ref{fig:sequence}; a
detailed investigation of individual features follows in
Sect.\,\ref{sect:FeH_magnetic}. The strengthening of the absorption
features towards later spectral types is clearly visible. Saturation
of the strongest lines becomes important after spectral type M5. In
cooler objects, even features that are weak in early M-type spectra
become comparable to the stronger lines, and eventually fill up the
gaps between individual lines. It is not immediately clear whether the
saturated spectra of the coolest objects are made up of the same
individual absorption features which are discernible in the early
M-stars. Our goal is to study the magnetic sensitivity of individual
features, which are most evident in spectra of slowly rotating early-
to mid-type M stars. In order to extrapolate this method to late-type
M stars and even to L-type dwarfs, we have to make sure that the weak
absorption features in the early-type M stars are due to the same FeH
lines as those in the strong absorption bands in the early L-type
dwarfs.

\clearpage\clearpage

\begin{figure*}
  \centering \leavevmode
  \includegraphics[width=\textwidth,bbllx=0,bblly=210,bburx=648,bbury=468]{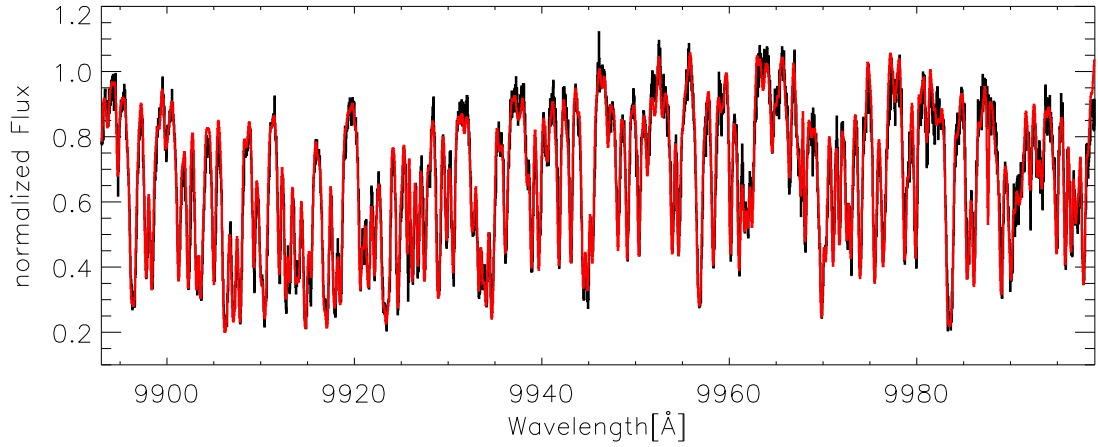}
\caption{\label{fig:VB10ampl}The FeH band in VB10 (M8, black), overplotted
  with a scaled version of the spectrum of GJ\,1227 (red; $\alpha =
  3.7$, cp. Eq.\,\ref{eq:cog}). No rotational broadening was applied.}
\end{figure*}

In Fig.\,\ref{fig:VB10ampl} we show the spectrum of VB\,10, an
ultracool dwarf of spectral type M8. The strongest FeH absorption
features are already severely saturated, while lines that were weak in
earlier-type M stars are strong in VB\,10. Thus, the difference
between strong and weak absorption features of FeH is much smaller
than it is in the early-type M dwarfs. In other words, a simple
scaling of the strength of absorption features observed in early-type
M dwarfs does not yield the absorption pattern observed in a late-type
M dwarf, because of spectral line saturation. We find, however, that
the pattern is well reproduced if one accounts for saturation, using a
scaling that is inspired by scaling optical depth:
\begin{equation}
   \label{eq:cog}
   S(\lambda) = 1 - C( 1 - A^\alpha).
\end{equation}
Here, $A(\lambda)$ is the normalized residual intensity at wavelength
$\lambda$ (the original spectrum with the continuum at normalized flux
unity). $\alpha$ is the optical depth scale factor which is applied to
the entire target spectrum. $C$ is a constant which controls the
maximum absorption depth due to saturation (this is necessary to match
the cores of strong lines, which do not go to zero, even for large
increases in optical depth), and $S(\lambda)$ is the amplified
absorption spectrum, weaker FeH absorption can also be calculated
using $\alpha < 1$. The red line in Fig.\,\ref{fig:VB10ampl} shows the
spectrum of the M4.5 dwarf GJ\,1227 amplified by a factor of $\alpha =
3.7$ using a saturation level of $C = 0.9$. The amplified spectrum of
GJ\,1227 matches the strong absorption features in the spectrum of
VB\,10 remarkably well.  It appears that the structure of FeH
absorption is identical in the M4.5 dwarf GJ\,1227 and in the cooler
M8 object VB\,10 -- one simply sees the effects of increasing
saturation of weaker lines in the cooler atmosphere.

\begin{figure}
  \plotone{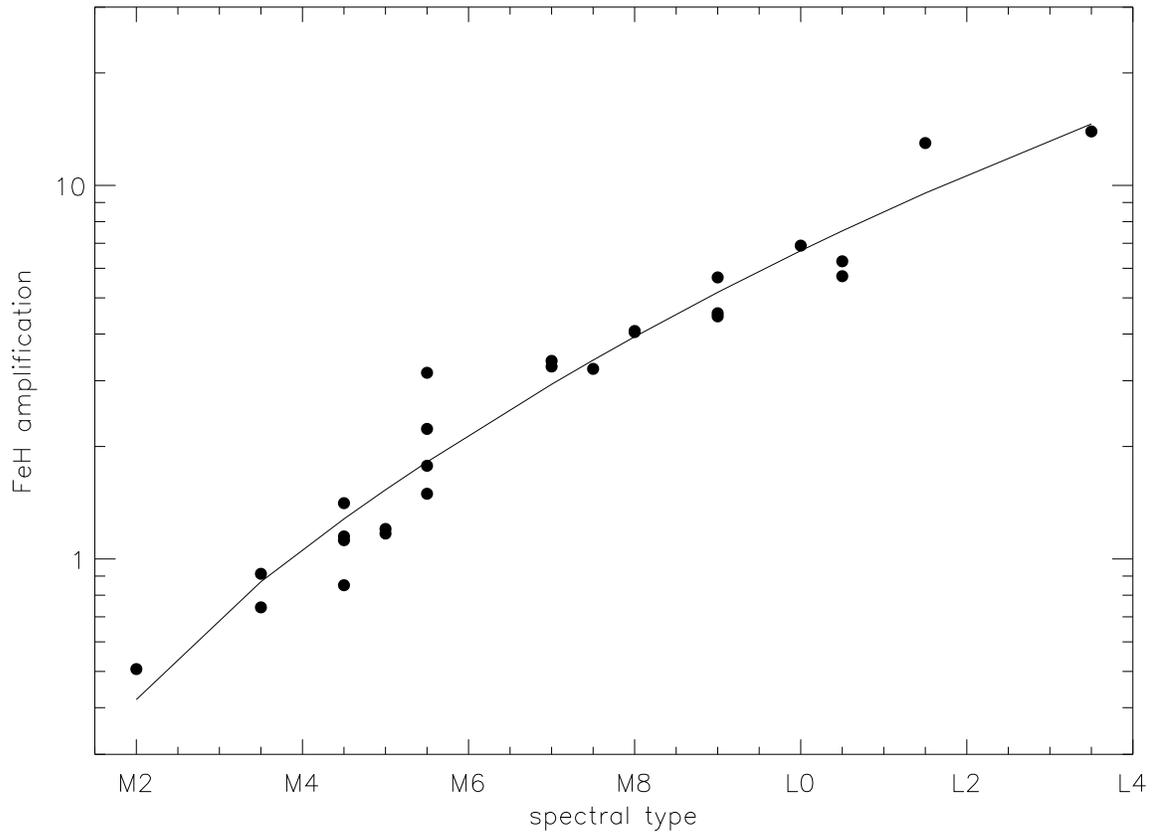}
  \caption{\label{fig:amplification}Scaling of the FeH 
    band in M and L-type objects; the amplification factor $\alpha$
    from Eq.\,\ref{eq:cog} is plotted as a function of spectral type.}
\end{figure}

We applied the same strategy to 25 objects with spectral types between
M2 and L4 and found that in all cases the scaled spectrum of GJ\,1227
matches the other FeH spectra very well. The values of $\alpha$
(Eq.\,\ref{eq:cog}) that we found for these objects are plotted
against their spectral types in Fig.\,\ref{fig:amplification}. We
confirm stronger FeH absorption towards later spectral types among M
dwarfs and at spectral type early L, which was shown by
\citet{McLean03} and \citet{Cushing05}.  This scaling relation is very
helpful when searching for Zeeman broadening in spectra of objects at
spectral types for which no reference spectra are available. We can
now compare them to scaled versions of spectra of stars with known
magnetic fields using the appropriate value of $\alpha$ for that
spectral type. It is also very useful for determining rotation
velocities (this has to be done first in order to look for Zeeman
broadening) in L-type dwarfs, which have no slowly rotating
counterparts they could be compared to. The rotation velocity can be
directly determined by comparing their FeH spectrum to a scaled
spectrum of a slowly rotating mid-type M dwarf. However, a
determination of rotation velocities in our ultracool objects is
beyond the scope of this paper and we will discuss this topic in more
detail in a later paper, in which we also discuss the 25 objects above
in more detail.

\clearpage

\clearpage\clearpage

\subsection{CrH absorption in the Wing-Ford band}

\begin{figure*}
  \centering \leavevmode
  \includegraphics[width=\textwidth,bbllx=0,bblly=210,bburx=648,bbury=468]{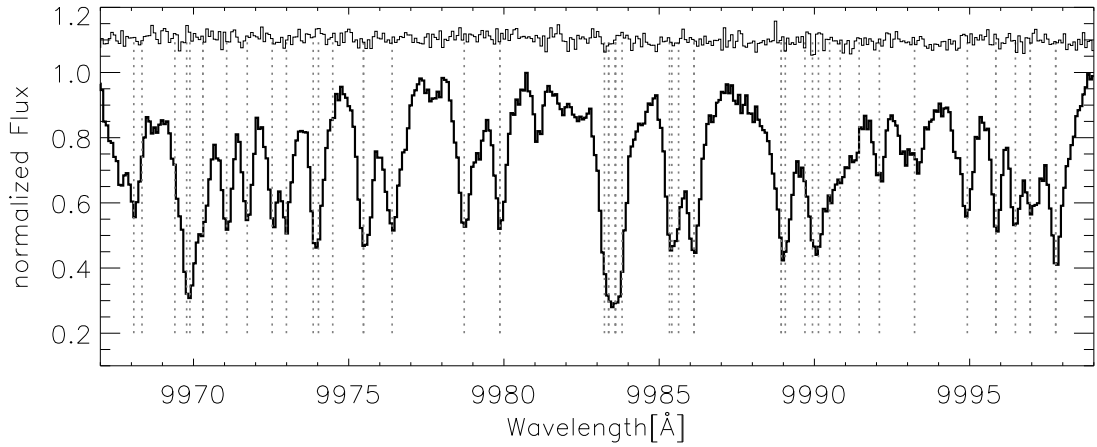}
  \caption{\label{fig:CrH}The spectrum of the M7 object LHS\,3003
    in the region where the CrH band is expected. Overplotted is a
    telluric reference spectrum of a B2IV star (with an offset of 0.1
    for better visibility). Positions of FeH lines are marked with
    dashed lines; we see no evidence for CrH absorption.}
\end{figure*}

A second absorption band that falls within the wavelength region of
our observation is the $0 - 1$ band of the $A\,^6\Sigma^+ -
X\,^6\Sigma^+$ system of CrH ($9969\,\mu$m). The detection of this
band has been claimed by \cite{Burrows02} who computed new line lists
and opacities.  \cite{Cushing03} question the identification of this
CrH bandhead since in their laboratory spectra they found strong FeH
features in this region as well. This is in fact important when
classifying L dwarfs on the basis of spectral indices
\citep{Kirkpatrick99}. From their low-resolution data,
\cite{Cushing03} could not determine to what extent CrH is present
around 9970\,\AA. We show our high resolution spectrum of the slowly
rotating M7 object LHS\,3003 in Fig.\,\ref{fig:CrH} together with a
spectrum of the B2IV star $\zeta$\,Cas as a telluric reference.
Positions of FeH lines from \cite{Dulick03} are indicated as dashed
lines.\footnote{We thank J.  Valenti for providing electronic versions
  of the FeH line data.}  Absorption features appear at every
predicted FeH position.  There is only one obvious absorption feature
at 9981.0\,\AA\ that is apparently not due to FeH, but this comparably
small feature does not have a significant influence on the overall
opacity. We thus conclude that no CrH absorption is visible in the
spectrum of this M7 object. At spectral type M7, absorption in this
wavelength interval is entirely caused by FeH. The structure of the
band does not change towards early-type L dwarfs (see
Fig.\,\ref{fig:sequence}), which means that even there no evidence for
CrH absorption was found.

\clearpage

\clearpage

\section{Magnetic sensitivity in the FeH band}
\label{sect:FeH_magnetic}

Magnetic field measurements utilize the space quantization of the
atomic angular momentum \textbf{\emph{J}} in a magnetic field; this is
called the Zeeman effect \citep[see for example][and references
therein]{Solanki91}. The sensitivity of atomic absorption lines to a
magnetic field is approximately proportional to the Land\'e factor
$g$, which is well known for many transitions. Magnetic splitting in
atomic lines can thus be calculated very precisely, and the recent
success of radiative transfer models in measuring magnetic fields in
cool dwarfs is summarized in \cite{Saar01, Valenti04}.

The molecular Zeeman effect is more complex than the atomic one. The
angular momentum vector has more quantization states due to nuclear
rotation, which has no dipole moment in the atomic case.  The coupling
of different angular momenta in the molecule -- nuclear rotation,
electronic orbit, and electron spin -- has to be taken into account.
This makes a calculation of the Land\'e-g factors more difficult than
in the atomic case. The coupling can be approximated by Hund's cases
(a) and (b) \cite[e.g.][]{Herzberg50}. A discussion of the molecular
Zeeman effect can be found in \cite{Crawford34}. The couplings of the
angular momentum of the $^4\Delta - ^4\Delta$ system of FeH is
intermediate between the limiting cases (a) and (b)
\citep{Phillips87}, but perturbation analysis has not yet been done
and spin-orbit coupling constants are still unknown.
\cite{Berdyugina02} show calculations for the Q branch of case (a),
the P- and R-branches are zero in this case (lines of the P-, Q- and
R-branches have $\Delta J = -1, 0, +1$, respectively.).  They expect
the perturbations to be comparable to the TiO~$\gamma$' system.  This
means that magnetic sensitivity in the P- and R-branches of FeH lines
would be high in transitions with very large values of $J$, and only
little around $J \la 15$. In the Q-branch, only the low-$J$
transitions are expected to be magnetically sensitive, since
perturbations are less important \citep{Berdyugina02}. However, the
intermediate coupling of angular momenta makes it difficult to make
precise calculations for Land\'e factors for this molecule, and
despite the efforts to understanding the FeH spectrum, its coupling
constants have not yet been measured and are still unknown.

In astronomical spectra, strong line broadening has been found in the
FeH lines of a sunspot spectrum \citep{Wallace98}. Those authors also
investigated the broadening of some lines as a function of $J$.
\cite{Berdyugina02} calculated the behavior of FeH lines for Hund's
case (a) and point out that perturbations appear to be similar to the
TiO~$\gamma'$ system.  They find that absolute values of $g_{\rm eff}$
of lines in the R- and P-branches will increase as $J$ increaes.
Qualitatively, this is what was observed by \citep{Wallace98}. A clear
demonstration of the potential of FeH as a tracer for magnetic fields
in cool stars has been given in \cite{Valenti01}. They show that some
Zeeman-sensitive lines in the spectrum of the active star AD~Leo (M3)
are much broader than in the inactive star GJ~725\,B (M3.5), while
insensitive lines have similar widths. It is obvious that FeH has an
excellent potential for measuring magnetic fields in cool dwarfs.
However, since coupling constants are still lacking, it is not yet
possible to measure magnetic fields by comparison to synthetic
spectra. A direct comparison to a sunspot spectrum is also quite
dangerous, since in the small spatial area observed, magnetic field
lines probably all point towards the same direction.  In contrast to
the situation in an integrated star spectrum, this means that certain
polarization components of the FeH lines may be missing, and the
behavior of a sunspot spectrum can significantly differ from the
spectrum of a star with a comparable magnetic field. However, a
sunspot spectrum could in principle be used to determine empirically
Land\'e--$g$ factors for the upper and lower levels of FeH lines.

\begin{figure}
\epsscale{.80}
  \plotone{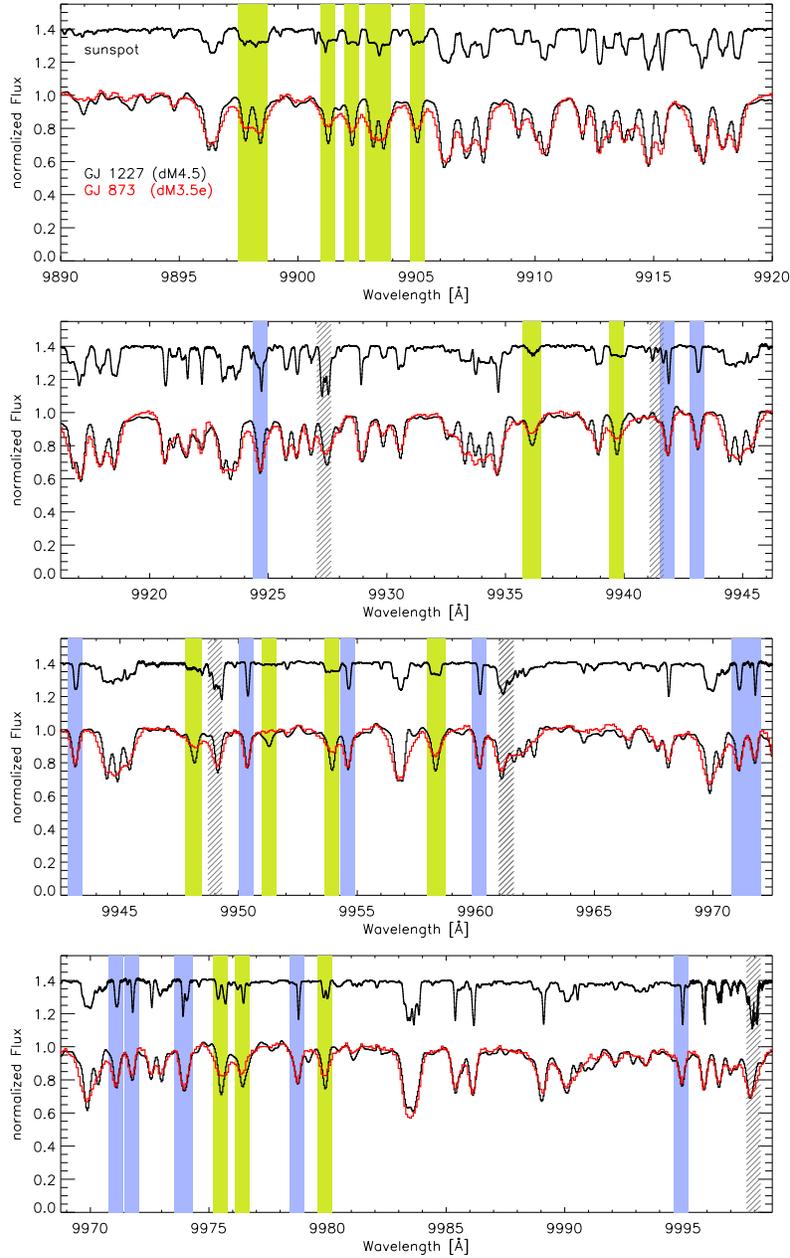}
  \caption{\label{fig:FeH}High resolution spectra of the inactive star 
    GJ\,1227 (M4.5, lower black line) and the active star Gl\,873
    (M3.5e). In order to compare the spectra, we artificially enhanced
    the absorption strength in Gl\,873 and spun up the spectrum of
    GJ\,1227 (see text). For comparison, the sunspot spectrum is
    overplotted with an offset. Magnetically insensitive lines are
    highlighted in blue, sensitive lines in green. Positions of atomic
    lines are marked as hatched regions.}
\end{figure}

Magnetic fields have been measured in early- and mid-type M dwarfs
using well-understood atomic lines. In these stars, the FeH band is
already prominent as well. For the technique we introduce here, we
compare a spectrum of an inactive star which presumably has no
measurable magnetic field, and one of an active star with a magnetic
field measured from atomic lines, to the spectrum in which we want to
detect magnetic broadening. The stars we use are given in
Table\,\ref{tab:mag_targets}, together with their X-ray and H$\alpha$
activity, and magnetic field measurements from \cite{JKV00} if
available. In the four panels of Fig.\,\ref{fig:FeH}, we show the FeH
region in our spectra of GJ\,1227 (black, inactive star, zero $Bf$)
and GJ\,873 (red, active star, $Bf = 3.9$\,kG). The spectral subtypes
are slightly different, and the FeH absorption in the cooler star
GJ\,1227 is a little stronger than it is in GJ\,873, which has a
slightly higher projected rotation velocity than the former. To make a
proper comparison, we amplified the absorption strength in the
displayed spectrum of GJ\,873 by a factor of $\alpha = 1.33$ using
Eq.\,\ref{eq:cog}, and we artificially broadened the spectrum of
GJ\,1227 according to a rotational velocity of 4\,km\,s$^{-1}$. Above
the two M-star spectra, we show the sunspot spectrum of
\cite{Wallace98}. The remaining differences in the two non-solar
spectra (Gl\,873 and GJ\,1227) plotted in Fig.\,\ref{fig:FeH} are due
to the different magnetic fields on the two stars.  The spectrum of
GJ\,1227 shows FeH lines without significant distortion by magnetic
fields; all individual absorption lines are narrow. The spectrum of
Gl\,873 exhibits the effect of magnetic line broadening in a star that
has a strong magnetic as well as a non-magnetic component (a filling
factor $f$ smaller than 1). The sunspot spectrum shows the extreme
case of magnetic broadening with no contribution of the non-magnetic
stellar surface outside the spots ($f = 1$) and minimal contribution
of Zeeman components near line center. This confirms the magnetic
nature of the differences between the M stars.

We identified two sets of absorption lines that show different
behavior in the magnetic and the non-magnetic cases. Eleven lines that
appear relatively insensitive to the Zeeman effect are listed in
Table\,\ref{tab:FeHconst}; two of them is are blends of two individual
lines (we list all lines that contribute to a blend and mark them in
the Tables). They are highlighted in blue in Fig.\,\ref{fig:FeH}.
Lines that are particularly sensitive to magnetic fields, i.e., lines
that appear significantly different in the two spectra of GJ\,1227 and
Gl\,873, are listed in Table\,\ref{tab:FeHvar} and are highlighted in
green in Fig.\,\ref{fig:FeH} (four of the features are blends of two
lines each). In this paper, we do not investigate the dependence in
the magnetic sensitivity of lines with different quantum numbers. Our
lists of magnetically insensitive and sensitive lines in
Tables\,\ref{tab:FeHconst} and \ref{tab:FeHvar} are not comprehensive.

Nevertheless, it easy to see in Table\,\ref{tab:FeHconst} that with
very few exceptions, the insensitive lines have $\Omega$ of 0.5 or
1.5, while the sensitive lines have $\Omega$ of 2.5 or 3.5, as
expected from the work of \cite{Wallace98}\footnote{We thank 
  J.\,Valenti for bringing this to our attention.}. This confirms the
strong dependence of $g_{\rm eff}$ on $\Omega$, the component of the
total electronic angular momentum along the internuclear axis.
Furthermore, all identified lines that are insensitive to Zeeman
broadening belong to the R-branch, most have quantum numbers $J < 15$.
On the other hand, very sensitive lines were found in all three (R-,
P-, and Q-) branches.  R-branch lines predominantly have very high
$J$-values ($J \sim 20$, but two lines have $J = 2.5$), and the two
Q-branch lines have very low $J$. Only very few P-branch lines are
identified, and only one ($J = 15.5$) is not part of a blend. We
conclude that magnetic line broading in the stellar spectra is in good
agreement with the results from \cite{Wallace98}, namely that magnetic
sensitivity strongly depends on the quantum number $\Omega$, a fact
that is not accounted for in a standard $^4\Delta-^4\Delta$
Hamiltonian. We also confirm that R-branch lines with very high
$J$-values and Q-branch lines with very low $J$-values show strong
magnetic sensitivity in qualitative agreement with the predictions
from \cite{Berdyugina02}.

A few atomic lines can be seen in this wavelength region as well. In
the spectral region of interest, the only atomic lines we expect in
the sunspot spectrum and our spectra of late-type M and early-type L
dwarfs are the Ti\,I lines with high oscillator strengths and low
excitation; four such lines are in the region between 9900 and
10\,000\,\AA. Two Na\,I lines may also appear. Line data from the VALD
database \citep{Kupka99} for the six atomic lines are listed in
Table\,\ref{tab:atomic}. Land\'e\,$g$ values are only known for the
Ti\,I lines. We mark atomic lines with hatched regions in
Fig.\,\ref{fig:FeH}.

The Wing-Ford band contains a large number of lines that are
individually resolved in the spectra of slow rotators.  Some of these
lines are rather insensitive to a magnetic field, while others show
Zeeman broadening that is comparable to rotational broadening on the
order of 10\,km\,s$^{-1}$. This has the effect of changing the
apparent strength of spectral features in a way that is more obvious
than the measurement of line broadening. Furthermore, the structure of
the FeH band does not vary between ultracool stars with different
temperatures so long as the saturation effect is taken into account.
Thus the Wing-Ford band provides one of the best opportunities to
investigate Zeeman broadening in cool stars and even in late-type M
and L dwarfs, at least so long as the signature is not washed out by
rapid rotation. The detectability of magnetic fields in cool rotating
objects is discussed in the next section.

\clearpage

\begin{deluxetable}{llrrc}
  \tablecaption{\label{tab:mag_targets} Targets for magnetic field measurements}
  \tablewidth{0pt}
  \tablehead{\colhead{Name} & \colhead{Sp}& \colhead{log($\frac{L_\mathrm{X}}{L_\mathrm{bol}}$)} & \colhead{log($\frac{L_\mathrm{H\alpha}}{L_\mathrm{bol}}$)}  & \colhead{$Bf$ [kG]}}
  \startdata
  Gl 729          & M3.5e & $ -3.50$ &          & 2.0\tablenotemark{1}\\
  Gl 873          & M3.5e & $ -3.07$ & $-$3.70  & 3.9\tablenotemark{1}\\
  GJ 1227         & M4.5  & $<-3.85$ & $<-5.0$  & \\
  \enddata
  \tablenotetext{1}{\citet{JKV00}}
\end{deluxetable}

\clearpage\clearpage

\begin{deluxetable}{lcrc}
  \tablecaption{\label{tab:FeHconst} Magnetically insensitive
    lines in Fig.\,\ref{fig:FeH}, these lines are highlighted in blue} \tablewidth{0pt} \tablehead{\colhead{$\lambda$ [\AA]} & Branch & $J$ & $\Omega$} 
\startdata
  9924.6374* &   R & 6.5  & 1.5\\
  9924.6374* &   R & 10.5 & 3.5\\
  \noalign{\medskip}
  9941.8298  &   R & 9.5  & 1.5\\
  9943.0934  &   R & 11.5 & 1.5\\
  9950.3384  &   R & 10.5 & 1.5\\
  9954.5883  &   R & 12.5 & 1.5\\
  9960.1308  &   R & 11.5 & 1.5\\
  9971.0688  &   R & 12.5 & 1.5\\
  9971.7306  &   R & 4.5  & 0.5\\
  \noalign{\medskip}
  9973.8550* &   R & 5.5  & 0.5\\
  9974.0250* &   R & 17.5 & 2.5\\
  \noalign{\medskip}
  9978.7207  &   R & 6.5  & 0.5\\
  9994.9243  &   R & 7.5  & 0.5\\
  \enddata 
  \tablenotetext{*}{The observed feature is a blend of the two
    neighbored lines.}
\end{deluxetable}

\clearpage

\begin{deluxetable}{lcrc}
  \tablecaption{\label{tab:FeHvar} Magnetically very sensitive lines in Fig.\,\ref{fig:FeH}, these lines are highlighted in green}
  \tablewidth{0pt}
  \tablehead{\colhead{$\lambda$ [\AA]} & Branch & $J$ & $\Omega$}
  \startdata
  9897.7737  &  R & 17.5 &  3.5\\
  9898.4006  &  R & 14.5 &  3.5\\
  9901.2588  &  R & 18.5 &  3.5\\
  9902.2715  &  R & 13.5 &  3.5\\
  9903.1523  &  R & 15.5 &  3.5\\ 
  9903.5932  &  R & 14.5 &  3.5\\
  9905.0208  &  R & 16.5 &  3.5\\
  \noalign{\medskip}
  9936.0393* &  R &  2.5 &  2.5\\
  9936.1883* &  R &  2.5 &  2.5\\
  \noalign{\medskip}
  9939.6853* &  P & 15.5 &  2.5-3.5\\
  9939.6853* &  P & 21.5 &  3.5\\
  \noalign{\medskip}
  9948.0461* &  P & 10.5 &  0.5-1.5\\
  9948.1340* &  R & 23.5 &  3.5\\
  \noalign{\medskip}
  9951.2732  &  P & 16.5 &  2.5-3.5\\
  9953.9115  &  R & 22.5 &  3.5\\
  \noalign{\medskip}
  9958.2533* &  R & 18.5 &  2.5\\
  9958.4052* &  P & 16.5 &  2.5-3.5\\
  \noalign{\medskip}
  9975.4756  &  Q &  2.5 &  2.5\\
  9976.3973  &  R &  2.5 &  0.5\\
  9979.8674  &  Q &  3.5 &  2.5\\
  \enddata
  \tablenotetext{*}{The observed feature is a blend of the two
    neighbored lines.}
\end{deluxetable}

\clearpage

\begin{deluxetable}{cccccc}
  \tablecaption{\label{tab:atomic} Atomic lines in ultracool stars$^*$}
  \tablewidth{0pt}
    \tablehead{\colhead{Ion} & \colhead{Wavelength [\AA]} &
    \colhead{J--J} & \colhead{log\,($gf$)} & \colhead{Energy levels
      [eV]} & \colhead{Land\'e $g$}}
  \startdata
   Ti I & 9927.35   &  4--3 & $-$1.580 & 1.8790  -- 3.1280 & 1.060\\
   Ti I & 9941.38   &  2--3 & $-$1.821 & 2.1600  -- 3.4070 & 1.550\\
   Ti I & 9949.00   &  1--2 & $-$1.778 & 2.1540  -- 3.3990 & 1.510\\
   Ti I & 9997.96   &  3--2 & $-$1.840 & 1.8730  -- 3.1130 & 0.810\\
   \noalign{\smallskip}
   Na I & 9961.26   &  2.5--3.5 & $-$0.820 & 3.6170  -- 4.8610\\
   Na I & 9961.31   &  1.5--2.5 & $-$0.980 & 3.6170  -- 4.8610\\
  \enddata
  \tablenotetext{*}{Data are from the Vienna Atomic Line Database \citep{Kupka99}}
\end{deluxetable}

\clearpage

\section{Detectability of Magnetic Fields on Ultracool Objects}
\label{sect:detectability}

Only very few atomic lines are sufficiently isolated and magnetically
sensitive to allow a detection of Zeeman broadening, and often only
one spectral line is investigated in studies of magnetic fields in
cool stars. In the wavelength region around 1\,$\mu$m, however, FeH
produces a number of magnetically sensitive and well-isolated lines.
The strong signal that is seen in the sensitive FeH lines allows
detection of pervasive magnetic fields even in the presence of
moderate rotation. In general, one can determine line broadening due
to rotation (and perhaps other mechanisms like turbulence) by
comparison of magnetically \emph{insensitive} lines to the same lines
in a reference star with known rotational velocity. The Zeeman signal
is analyzed by comparing the shape of magnetically \emph{sensitive}
lines between the target spectrum and the reference spectrum. If the
Zeeman splitting is small compared to the intrinsic line width, one
can only measure the magnetic flux ($Bf$); disentangling the filling
factor ($f$) from the magnetic field strength ($B$) is rare in stellar
cases \citep{VMB95}.  If the magnetic flux in the reference spectrum
is known, one can place limits on the flux in the target spectrum.
Having reference spectra for several different flux levels allows
refinement of the measurement.

\subsection{Magnetic Measurement by Line Ratios}
\label{sect:Bfratio}

One method of obtaining the magnetic signal is to employ line-depth
ratios between magnetically sensitive and insensitive lines.  The
expectation is that the line-depth ratio between a Zeeman-sensitive
and Zeeman-insensitive line depends monotonically on the magnetic
field, since the insensitive line remains unaffected while the
sensitive line becomes shallower in a stronger field. Such a ratio can
be calibrated using reference spectra with known magnetic field
strengths. 

The two lines should be chosen in close proximity to each other, so
that differential errors in continuum placement are less of a concern.
Comparing line depths requires the spectra to be observed or broadened
to a common resolution. Since rotation and resolution essentially have
the same effect on the line widths, the limiting resolution of this
method is the one that corresponds to the maximum rotation velocity
for which the method can be applied.

We used line-depth ratios to investigate the range of parameters for 
which a detection of magnetic fields through the FeH Zeeman signal is
feasible. We identified four ratios of neighboring absorption
features that are particularly useful to measure the magnetic flux -- 
two for slow rotators ($v\,\sin{i} \la
20$\,km\,s$^{-1}$) and two for rapid rotators ($v\,\sin{i} \ga
15$\,km\,s$^{-1}$).  The lines used for these ratios are given in
Table\,\ref{tab:ratios}. For the rapid rotators, the positions are not
centered on a physical absorption line, but rather on a feature that
is a blend of several lines at that rotation velocity. The Zeeman
sensitivity of the blended lines is what we test. In order to 
investigate the detectability of the Zeeman signal 
at a given rotation velocity and given spectral type (which has a 
certain level of saturation), we used the template spectra of active
and inactive stars that we have shown in Fig.\,\ref{fig:FeH}, namely
spectra of GJ\,1227 (zero field) and Gl\,873 ($Bf \sim 3.9$\,kG). To
probe whether Zeeman broadening due to a field that is weaker than
$\sim4$\,kG is detectable, we constructed artificial spectra with
intermediate magnetic broadening by a linear interpolation of the
observed reference spectra using
\begin{equation}
\label{eq:S}
S_{\rm new} = pS_{\rm Gl\,873} + (1-p) S_{\rm GJ\,1227},
\end{equation}
with $S$ the spectra and $p = Bf/(3.9$\,kG).  This assumes that
magnetic broadening is a linear function of $Bf$, which may well not
be the case. Especially in the comparison to other stars, a different
fraction of surface coverage $f$ will change the structure of line
broadening. However, without any knowledge of $f$ and without a
detailed modeling of individual Zeeman components, this assumption is
probably the best we can do from the observational side.  So long as
molecular constants are missing, no definite statement about the
individual parameters $B$ and $f$ can be made (and even then they may
be difficult to disentangle). For now, we have to accept this as a
systematic uncertainty of our method.

\clearpage

\begin{deluxetable}{lcc}
  \tablecaption{\label{tab:ratios} Pairs of magnetically insensitive
    and magnetically sensitive features used in either slow
    ($v\,\sin{i} \la 20$\,km\,s$^{-1}$) or rapid ($v\,\sin{i} \ga
    15$\,km\,s$^{-1}$) rotators. The ratio between the depths of two
    features can be used as a proxy of the magnetic field (see text).}
  \tablewidth{0pt}
  \tablehead{& \colhead{Insensitive Line}& \colhead{Sensitive Line}\\
    &  $\lambda$ [\AA] & $\lambda$ [\AA]} \startdata
  \multicolumn{3}{c}{Slow Rotators}\\
  \noalign{\smallskip}
  1. & 9949.11 & 9948.13 \\
  \noalign{\smallskip}
  2. & 9956.77 & 9958.25 \\
  \noalign{\smallskip}
  \multicolumn{3}{c}{Rapid Rotators}\\
  \noalign{\smallskip}
  1. & 9948.7 & 9950.4 \\
  2. & 9979.8 & 9978.8 \\
  \enddata
\end{deluxetable}

\clearpage

\begin{figure}
  \centering \leavevmode
  \includegraphics[width=\textwidth,bbllx=0,bblly=130,bburx=568,bbury=789]{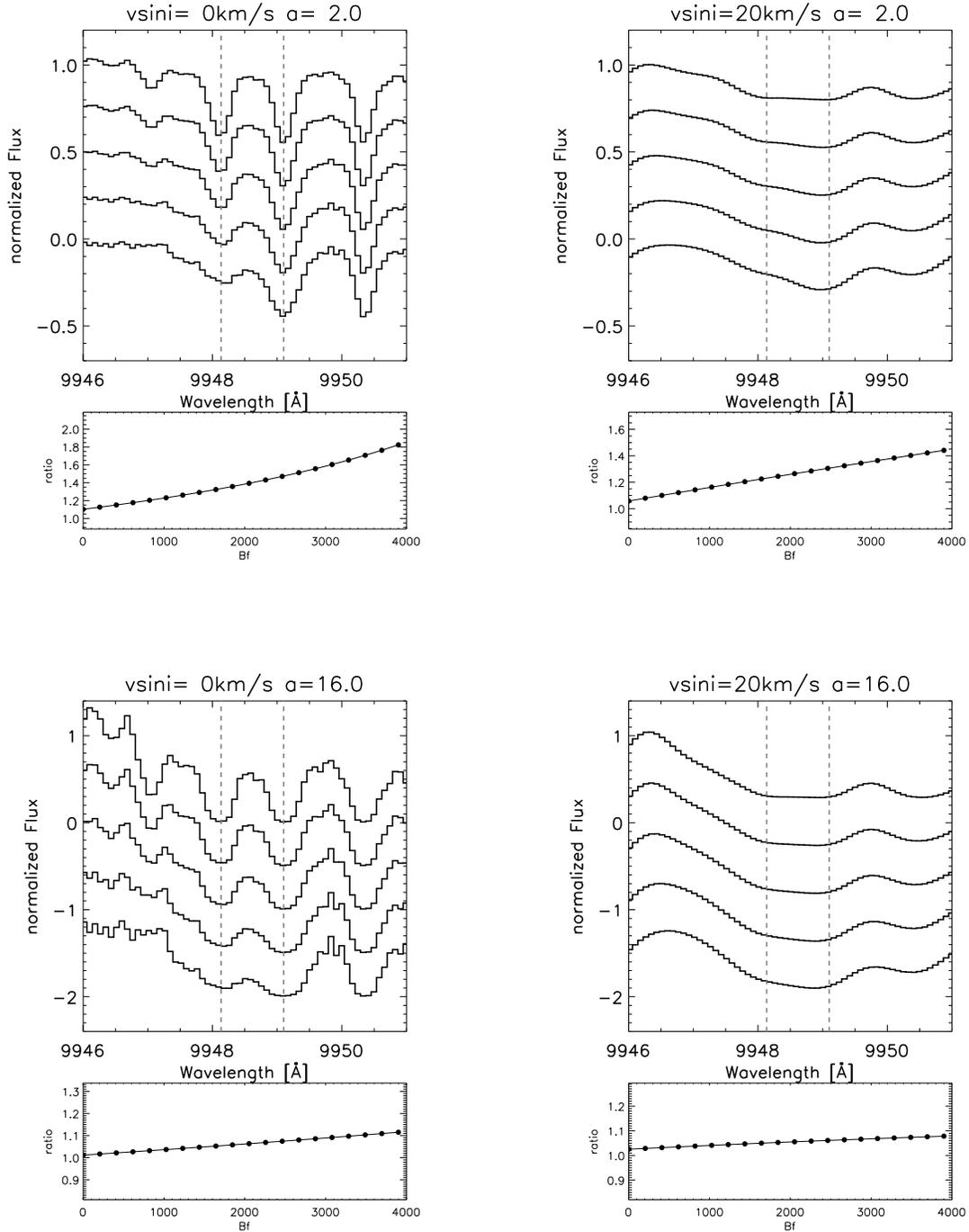}
  \caption{\label{fig:ratio_lines}Behavior of the first line pair useful 
    in slow rotators. Left panel shows the case of no rotation, in the
    right panel the spectra are spun up to $v\,\sin{i} =
    20$\,km\,s$^{-1}$.  The upper panel shows the case of
    non-saturated lines ($\alpha = 2, \sim$M6), in the lower panel the
    FeH band is heavily saturated ($\alpha=16, \sim$L4). In each of
    the four examples, the magnetic field increases from top to bottom
    in the spectra.  The ratio between the depths of the two
    absorption features, that are marked with dashed lines, are
    plotted as a function of $Bf = p\cdot3.9$\,kG in the small plot
    below each figure (cp Eq.\,\ref{eq:S}).}
\end{figure}

In Fig.\,\ref{fig:ratio_lines}, we show the predicted behavior of one
of the four line ratios of Table\,\ref{tab:ratios} with varying
magnetic field at different rotation velocities and FeH absorption
strength.  The sensitive and insensitive features are marked with
dashed lines.  In the upper row we show the ratio as observed in
objects of spectral type M6, i.e., $\alpha =2$. In the lower panel,
FeH absorption is saturated and corresponds to a spectral type of L4
($\alpha =16$). The left column shows spectra in slowly rotating
objects, the spectra in the right column are spun up to $v\,\sin{i} =
20$\,km\,s$^{-1}$. In each panel, five spectra of different ``magnetic
flux'' ($p \sim Bf$) are shown as an example -- the top spectra have
$p=0$ (pure GJ\,1227), the bottom have $p=1$ (pure Gl\,873), and the
intermediate spectra are linear interpolations of these two cases. In
the small figures below each example, we show the ratio of the depths
of the magnetically insensitive features to the sensitive features as
a function of $Bf = p\cdot3.9$\,kG. It is shown for more points than
plotted as examples.

The upper left panel of Fig.\,\ref{fig:ratio_lines} shows how the
magnetically sensitive line (left marked line) is washed out with
growing magnetic field strength while the insensitive line is only
marginally affected by the field. The ratio of the two lines, shown in
the small figure below, changes from about one in the non-magnetic
case to a value of two at $Bf \sim 4$\,kG. Such a difference is easily
detectable even in data of low SNR. It is no surprise that the two
entirely separated lines show monotonic behavior in their interpolated
ratio. On the other hand, at faster rotation or in saturated lines,
this behavior may be different since blending and saturation wash out
the differences between the two measured features. This is shown in
the other three panels of Fig.\,\ref{fig:ratio_lines}. In none of
them, the signal is as clear as in the pure case for no rotation
without saturation, although the line ratios are still growing with
larger field strengths. The amplitude is much smaller than in the pure
example.

\begin{figure}
  \plotone{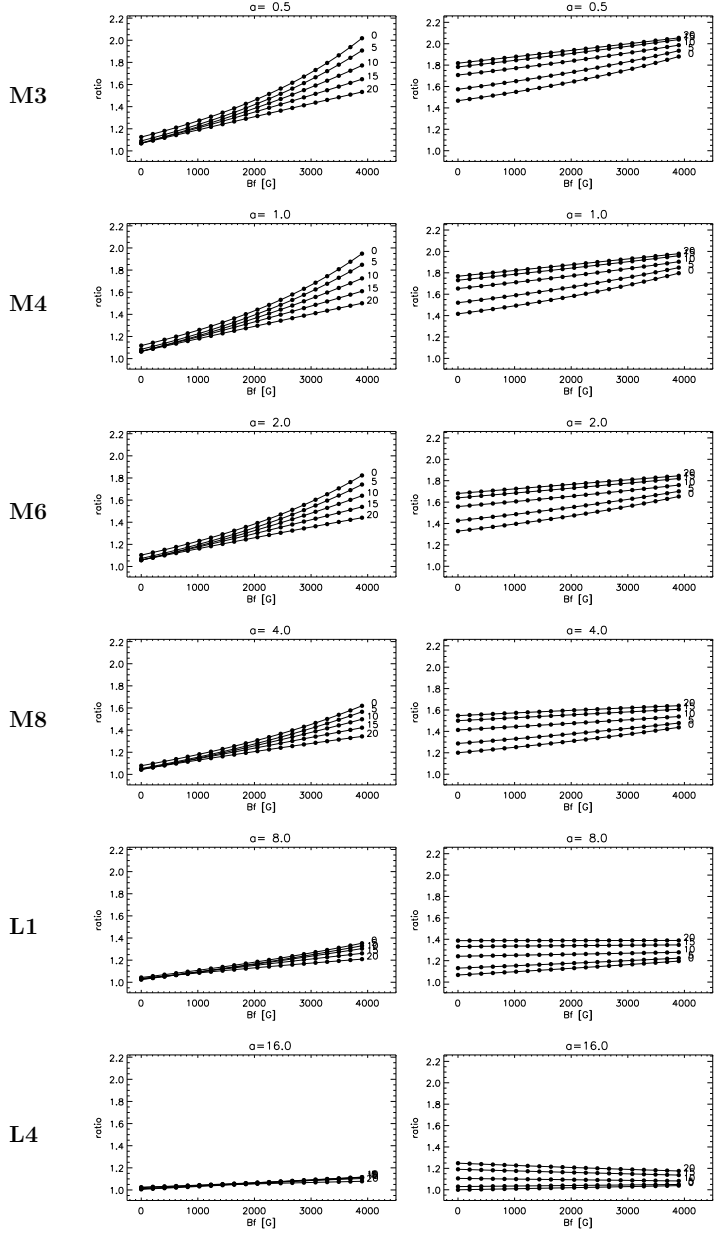}
  \caption{\label{fig:ratios_slow}Depth ratios for slow rotatos; left panel: 
    first ratio; right panel: second ratio. From top to bottom the
    rows show the results for different absorption strengths of FeH,
    the approximate spectral type is marked at the left side. In each
    plot, results for different values with 0\,km\,s$^{-1} \le
    v\,\sin{i} \le 20$\,km\,s$^{-1}$ are given as annotated in the
    figures.}
\end{figure}

\begin{figure}
  \plotone{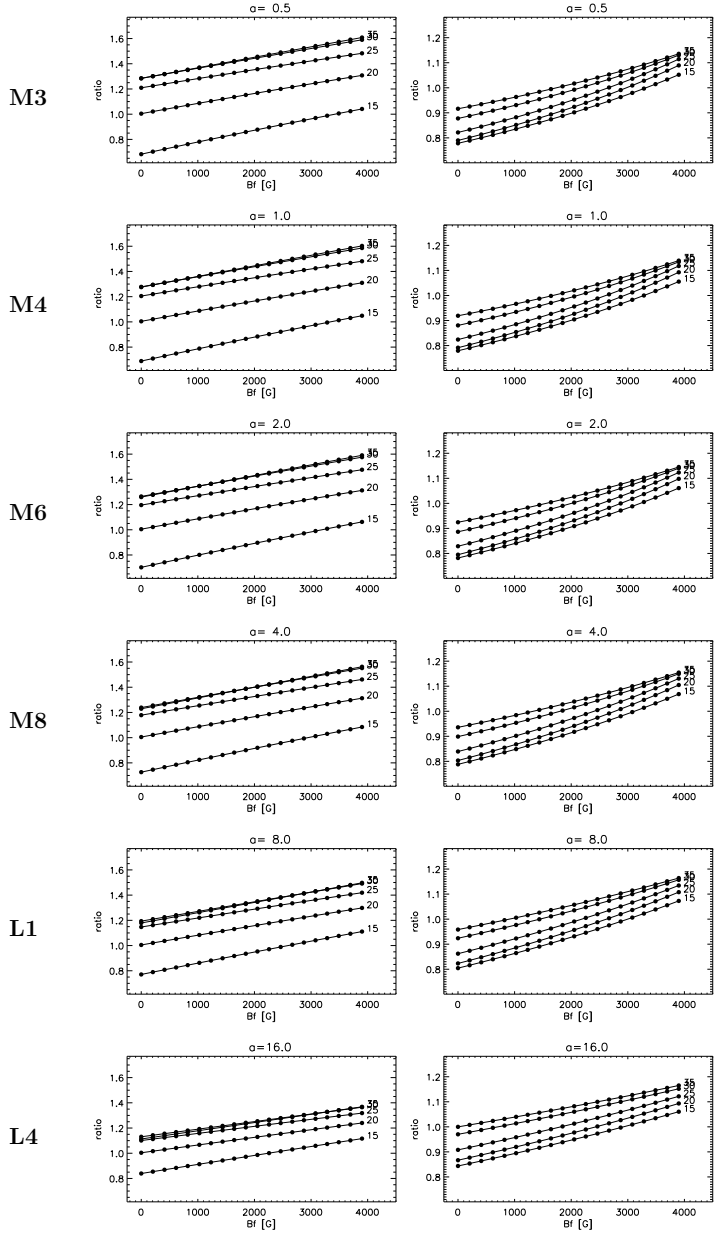}
  \caption{\label{fig:ratios_rapid}As Fig.\,\ref{fig:ratios_slow} but for the 
    features used in rapid rotators (15\,km\,s$^{-1} \le v\,\sin{i} \le
    35$\,km\,s$^{-1}$).}
\end{figure}

In Table\,\ref{tab:ratios}, we have introduced two line ratios that
can be used for slow rotators, and two line ratios that are better
suited in more rapidly rotating stars. The behavior of all four ratios
with varying field strengths are shown for different rotational
velocities and different FeH band strengths $\alpha$ (different
spectral types) in Figs.\,\ref{fig:ratios_slow} and
\ref{fig:ratios_rapid}; Fig.\,\ref{fig:ratios_slow} shows the results
for the ratios suited for slow rotators, Fig.\,\ref{fig:ratios_rapid}
shows the ones for rapid rotators. For each case, the left panel shows
the first ratio, the right panel shows the second ratio as given in
Table\,\ref{tab:ratios}. For each ratio, six plots are shown in the
six rows of Figs.\,\ref{fig:ratios_slow} and \ref{fig:ratios_rapid}.
They correspond to six different values of line depth
(Eq.\,\ref{eq:cog}), $\alpha\,=\,[0.5, 1, 2, 4, 8, 16]$, i.e.
approximately to the spectral types M3, M4, M6, M8, L1 and L4
(Fig.\,\ref{fig:amplification}). For the slow rotators, we calculated
the ratios for projected rotation velocities of $v\,\sin{i} = [0, 5,
10, 15, 20]$\,km\,s$^{-1}$, for rapid rotators we show $v\,\sin{i} =
[15, 20, 25, 30, 35]$\,km\,s$^{-1}$, the values of $v\,\sin{i}$ are
annotated in the plots.

Most of the ratios shown in Figs.\,\ref{fig:ratios_slow} and
\ref{fig:ratios_rapid} show a monotonic dependence in $p \sim Bf$, and
in many cases magnetic broadening has a detectable influence on these
ratios.  Thus, magnetic broadening can in principle be detected in
objects of spectral type as late as L4 and rotating as rapidly as
$v\,\sin{i} = 35$\,km\,s$^{-1}$. In practice, the limiting factors of
such a measurement are the signal-to-noise ratio (SNR) and the
determination of the continuum. If strong magnetic broadening in a
rapidly rotating object of spectral L4 changes a measured ratio from
1.0 to 1.15, this means that the ratio has to be measurable with such
accuracy. In this example, a SNR of 20 would suffice to detect
differences between the extreme cases of a field as strong as in
Gl\,873 ($Bf \approx 4$\,kG) and the case of no field. This SNR
includes the placement of the continuum, which is especially delicate
in rapid rotators where the continuum is no longer visible between
individual lines of FeH. This means that the method is more severely
limited by the data quality we can achieve, and especially by the
accuracy of the rotational velocity one can derive. One can imagine
developing more sophisticated techniques to deal with these difficulty
cases in the future.

We have obtained a spectrum of another M-dwarf for which the magnetic
field strength was measured from atomic lines. The magnetic flux in
the M3.5e star Gl\,729 is $Bf \approx 2.0$\,kG \citep{JKV00}, which is
intermediate between the flux found in Gl\,873 ($Bf \approx 3.9$\,kG)
and the zero flux in GJ\,1227. We now use its spectrum to check the
accuracy of measuring magnetic fields using the line ratios for slow
rotators introduced above. For M3.5e the saturation correction factor
is about $\alpha = 0.9$ (Eq.\,\ref{eq:cog}).  The projected rotational
velocity does not exceed that of GJ\,1227, which has a $v\,\sin{i}$
below 2.3\,km\,s$^{-1}$ \citep{Delfosse98}.  Slow rotation in Gl\,729
has also been found by \citet[] [$v\,\sin{i} =
3.5$\,km\,s$^{-1}$]{JKV96}.

The ratios most appropriate for Gl\,729 are hence the ones for slow
rotators, namely the line pairs around 9949\,\AA\ and 9957\,\AA. We
measured the ratios between the line depths $d$ in these pairs and
find $d$(9949.11\AA)/$d$(9948.13\AA) = $1.54 \pm 0.08$ and
$d$(9956.77\AA)/$d$(9958.25\AA) = $1.61 \pm 0.06$. The uncertainties
are for a SNR of 100, which is our estimate for the data quality
including the determination of the continuum. From these ratios, we
can now estimate the value of $Bf$ from the predicted ratios taken
from the interpolations in Sect.\,\ref{sect:detectability}.  We have
to apply this to the plots in the second top row ($\alpha = 1$) of
Fig.\,\ref{fig:ratios_slow} at the given rotation velocity (we used
the region 0--5\,km\,s$^{-1}$).  The result from the first line ratio
is $Bf = (2.45 \pm 0.45)$\,kG, the second ratio gives $Bf = (1.45 \pm
0.80)$\,kG. Both values are consistent with $Bf \approx 2.0$\,kG as
measured from atomic Fe lines, with an error of less than a kilogauss.
Furthermore, both values are inconsistent with zero magnetic flux and
are also inconsistent with a flux $Bf$ as strong as found on Gl\,873
($Bf \approx 4$\,kG).  This means, that with the line ratio method in
FeH the magnetic flux in Gl\,729 can be distinguished from zero flux
and from magnetic flux as strong as $Bf = 4$\,kG.

\clearpage

\clearpage

\clearpage\clearpage

\clearpage


\subsection{Magnetic Measurement by $\chi^2$ fitting}
\label{sect:Bfchisq}

Measuring the magnetic broadening through individual line ratios has a
number of weaknesses. Most importantly, line ratios are measured from
only a few pixels; in slow rotators the depth of a line may be
governed by only one pixel.  This means that this method is very prone
to noise.  Furthermore, the calculation of the line depth seriously
depends on the correct placement of the continuum, which may be
difficult to determine, particularly in objects of later spectral type
that have saturated FeH bands and in rapid rotators. Another caveat to
the line-ratio method is that the rotation velocity and FeH absorption
strength have to be known in order to translate a ratio to a magnetic
field strength. It would be much better to determine all unknown
values from a single fit that includes both magnetically sensitive and
insensitive lines.

As a second method of measuring the magnetic flux $Bf$ from FeH lines,
we construct an artificial spectrum from reference spectra of an
inactive star and an active star with known magnetic fields
(Eq.\,\ref{eq:S}), and search for the best fit to the spectrum of the
object in terms of $\chi^2$-fitting. The construction of the spectrum
is done by linear interpolation as before. In a first step, we match
their continua by simple scaling. Then we search for the best fit to
reproduce the target spectrum. The parameters of our fit are FeH
saturation strength $\alpha$, projected rotational velocity
$v\,\sin{i}$ and magnetic flux $Bf$. With this strategy, all three
parameters can be determined simultaneously, and in fact much more
accurately, than obtaining them individually. Since lines with
different dependencies on the magnetic field are involved, and since
the fit is over a significant wavelength region, the result is
generally more robust than a determination from one or two line ratios
can be. However, any technique that involves $\chi^2$-fitting entirely
relies on the assumption that a solution exists that can reproduce the
data, and that deviations between the data and the fit are essentially
due to noise. As soon as systematic differences become important in
the calculation of the deviation between data and fit, the results of
this technique are difficult to quantify. Thus, it is important to
check the reliability of the best fit, i.e., whether it reproduces all
significant features -- in particular both the magnetically sensitive
and insensitive features.  If a good fit can be achieved in a broad
wavelength region incorporating a number of absoprtion features, this
technique will be very sensitive to the magnetic field in terms of
$Bf$. Furthermore, if a model can accurately reproduce magnetically
sensitive \emph{and} insensitive lines, there will be little room for
systematic problems during the construction of the fit.  They should
produce significant deviations at least in some of the different
features.

\begin{figure}
  \plotone{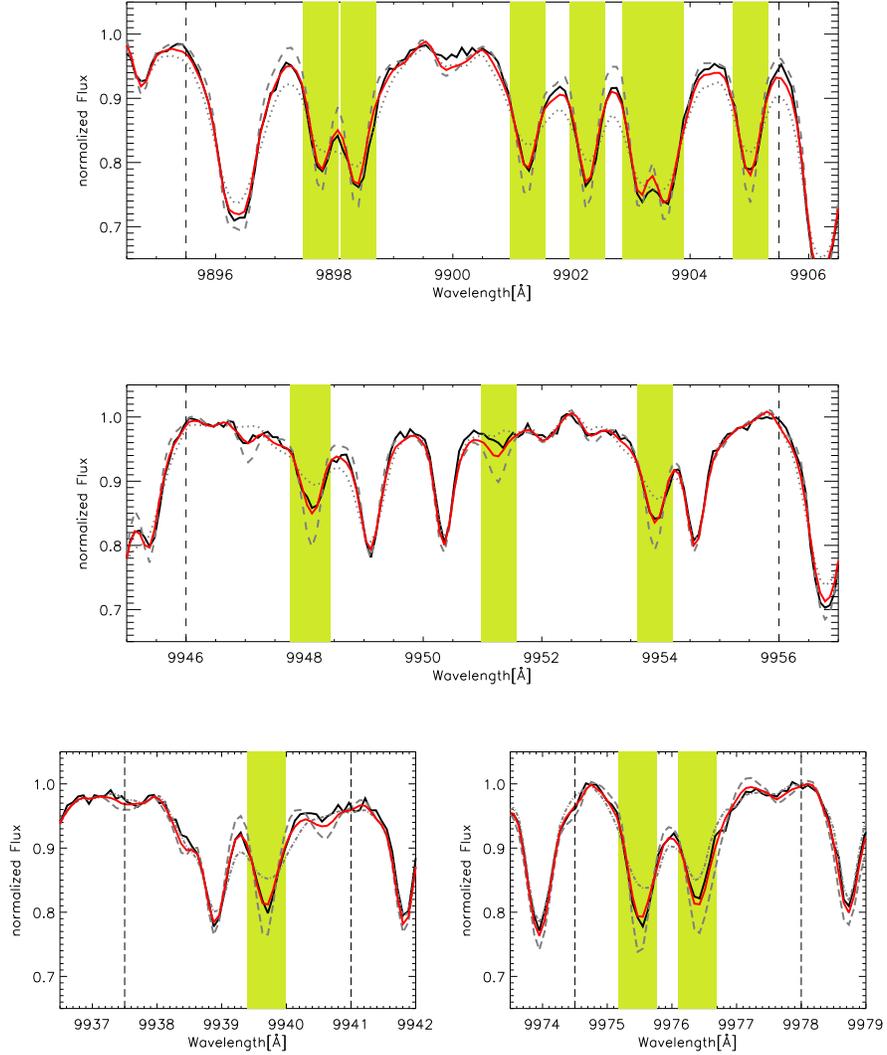}
  \caption{\label{fig:Gl729fit}Fit (red line) of a linear interpolation between the
    spectra of GJ\,1227 (grey dashed line) and Gl\,873 (grey dotted
    line) to the spectrum of Gl\,729 (black line). $\chi^2$ is
    calculated from the regions between the vertical dashed lines. The
    best fit is achievd for $Bf = 2.0$\,kG, i.e. $\sim 50\%$ GJ\,1227
    and $\sim 50\%$ Gl\,873.}
\end{figure}

As an example, we search for the best fit to our spectrum of Gl\,729
using an interpolation between the active star Gl\,873 and the
inactive star GJ\,1227 as done above. We identified four spectral
regions that are particularly useful for such a fit in the sense that
they contain a number of isolated magnetically sensitive and
insensitive lines. The parameters of the stars are given in the
previous section and in Table\,\ref{tab:mag_targets}.  We show the
best fit that we achieved in the four panels of
Fig.\,\ref{fig:Gl729fit}; the regions incorporated in the fitting
procedure are bracketed with dashed lines.  FeH lines that are
particularly sensitive to Zeeman broadening are marked in green. The
spectrum of the inactive star GJ\,1227 (zero $Bf$) is shown as a
dashed grey line, the spectrum of the active star Gl\,873 ($Bf \approx
3.9$\,kG) as a dotted grey line. Our observed spectrum of Gl\,729 is
plotted in black, it always falls between the two reference spectra.
The red line indicates our best fit, it is an interpolation of the two
reference spectra with a fraction of 54\,\% of the active star and
46\,\% of the inactive star. It provides a remarkably good fit over
the whole wavelength region. In particular, the fit can reproduce the
data of Gl\,729 in all lines that are magnetically sensitive, i.e.
where the interpolation lies between the two template spectra.
Assuming a linear dependence between line broadening and net magnetic
field, one derives a value of $Bf \approx 2.0$\,kG. We estimate the
uncertainty of the fit by the limits in $Bf$ at which $\chi_{\nu}^2 =
\chi_{\nu, \rm min}^2 + 1$ assuming $\chi_{\nu, \rm min}^2 = 1$, which
is the case for a SNR of 90.\footnote{We note that a measured SNR
  would be more appropriate.  However, the SNR in this spectral region
  is difficult to measure due to the many small features in the
  spectrum, SNR~=~90 is consistent with the data.} In other words, we
are looking for the values of $Bf$ that yield $\chi_{\nu}^2 = 2$.
This yields an an uncertainty of 1.3\,kG, i.e., $Bf \approx (2.0 \pm
1.3)$\,kG.

\clearpage\clearpage

\subsection{Line ratios vs. data fitting}

We have shown two ways of obtaining an estimate of the magnetic flux
from Zeeman modification of FeH spectra of cool stars.  Both methods
rely on the assumption that the spectrum of a star with intermediate
magnetic flux can be interpolated between two spectra, one with very
weak or zero flux and another with strong flux.  Although such an
interpolation is not necessarily linear, at this level of accuracy it
probably is a reasonable approximation. In practice, the limiting
factor of such a measurement is the SNR of the data. This includes the
determination of the continuum, which is problematic in spectra of the
cooler objects.

Fitting the interpolation of the two template spectra to the data has
the advantage that a large part of the spectral region is incorporated
and the result is based on a larger number of tracers while anchored
in the magnetically insensitive lines. This also yields very precise
values of the rotation velocity. However, especially in the cooler
stars where FeH is strongly saturated, and in rapid rotators
($v\,\sin{i} \ga 15$\,km\,s$^{-1}$), it will be difficult to reach the
desired fit quality, mainly because the continuum determination
becomes difficult. In this case the value of $\chi^2$ becomes
meaningless, since no reasonable fit can be achieved at all. Here the
line ratios can be more useful since they depend only marginally on
the correct placement of the continuum; they focus on a very narrow
spectral region. The fact that the errors we get from line ratios are
smaller than the one from the $\chi^2$-fit is because the formal
errors from the line ratio technique do not account for systematic
differences between the template and the target spectrum. The true
uncertainty including all systematic uncertainties would be larger --
these uncertainties are accounted for in the $\chi^2$-technique. The
two methods introduced in Sects.\,\ref{sect:Bfratio} and
\ref{sect:Bfchisq} are complementary and we suggest to derive the
value of a magnetic field from a direct fit and to check the result
using the line ratios. The $\chi^2$ method may not be applicable in
some cases where line ratios could still provide a reasonable field
estimate.

\section{Summary}
\label{sect:Summary}

We have analysed high resolution spectra of the FeH Wing-Ford band
near 1\,$\mu$m in very cool stars. The comparison of absorption
features to FeH line lists showed that all significant absorption
features around 9980\,\AA\ can be assigned to lines of FeH.  We did
not find extra absorption due to CrH where its 0--1 band was expected.
The depth of the FeH absorption lines scales inversely with stellar
temperature similar to a ``curve of growth'' analysis in spectral
lines.  This makes it possible to predict the spectrum of a
rapidly-rotating ultracool object on the basis of spectra of
slowly-rotating cool stars, permitting analysis of observed spectra in
the late M or L regime.  We found that FeH displays a number of
spectral lines that are individually resolved and intrinsically
narrow, and many of them show significant Land\'e-g factors. This is a
crucial advantage of the FeH Wing-Ford band compared with other
molecular bands, whose lines are easily blended by rapid rotation, and
which usually have very few magnetically sensitive lines strong enough
to be observed. Thus, the lines of the Wing-Ford band provide a unique
opportunity to investigate Zeeman broadening, as has been done
successfully for decades using atomic lines in hotter stars. It offers
the opportunity to extend such studies even to L-type dwarfs, which
has not otherwise been possible since slowly rotating L-dwarfs are
very rare.

We investigated individual spectral features of the two M-dwarfs
GJ\,1227 and Gl\,873; the latter is a known magnetically active star
with a magnetic flux of $Bf \approx 3.9$\,kG, while the former is
magnetically inactive and does not have a detectable magnetic field.
Comparison of the two spectra revealed a number of FeH lines that are
particularly sensitive to magnetic broadening -- such lines are
significantly broader in the magnetically active stars than in the
inactive one. Other lines show the same shape in both stars; these
lines are not sensitive to magnetic fields and demonstrate that the
extra broadening in the sensitive lines cannot be due to rotation or
other broadening mechanisms that are universal to all absorption
lines.

The molecular constants required for a full calculation of Zeeman
broadening in lines of molecular FeH are still unknown, hence Zeeman
signatures cannot yet be calculated as a function of magnetic field.
In order to measure a magnetic field in ultracool objects, we use
spectra of active and inactive stars with known magnetic fields to
calibrate the extra broadening seen in lines that are magnetically
sensitive.  An empirical investigation of four line-depth ratios between
magnetically sensitive and magnetically insensitive lines reveals
that in principle the effects of Zeeman broadening are detectable in
objects of spectral type late M and even in the L regime. Furthermore,
magnetic broadening produces signatures strong enough to appear in stars
rotating as rapidly as $v\,\sin{i} \approx 30$\,km\,s$^{-1}$, although
it may be difficult to reach the desired SNR for such a detection,
especially in the coolest objects.

In order to test the reliability of the proposed method, we tried to
redetermine the magnetic flux of the M3.5e star Gl\,729, which is
known from measurements in atomic lines. First, we used the line
ratios that we calibrated at the spectra of GJ\,1227 and Gl\,873, and
secondly, we fitted an interpolation of these two spectra to the data
of Gl\,729. Both methods yielded values of $Bf$ that are in reasonable
agreement with the field measured in the atomic lines. Most importantly,
for both strategies the uncertainties are small enough to
significantly exclude the assumption that Gl\,729 has no magnetic
field, or that it has a field that is as strong as Gl\,873. The
precision of this method seems to be of the order of 1\,kG.

The interpolation of FeH Zeeman signatures from two reference spectra
assumes that the magnetic flux signal is a linear function in $Bf$. This
assumption is a big simplification of the physical situation, and thus
implies that a precise measurement of $Bf$ (and especially a separation
of $B$ and $f$) cannot yet be expected from this method.  Clearly, a full
calculation of the Zeeman splitting in FeH lines would yield much more
accuracy in the determination of magnetic fields in cool stars. It is not
clear to what extent we could then separate the effects of $B$ and $f$.
Nevertheless, until the calculation of molecular Zeeman splitting in
FeH becomes achievable, the \emph{detection} of a magnetic field is by
itself very valuable. We can further get a crude idea of what the 
value of the magnetic flux actually is, and compare stars with broadly
different fields. This method provides the first chance of measuring
magnetic fields in ultra low-mass stars and brown dwarfs.

\acknowledgments It is a pleasure to thank J. Valenti, S. Solanki and
S. Berdyugina for enlightening discussions on the magnetic sensitivity
of the FeH molecule. We especially thank J. Valenti for his help with
the line data. This work is based on observations obtained from the
W.M. Keck Observatory, which is operated as a scientific partnership
among the California Institute of Technology, the University of
California and the National Aeronautics and Space Administration. We
would like to acknowledge the great cultural significance of Mauna Kea
for native Hawaiians and express our gratitude for permission to
observe from atop this mountain. GB thanks the NSF for grant support
through AST00-98468. AR has received research funding from the
European Commission's Sixth Framework Programme as an Outgoing
International Fellow (MOIF-CT-2004-002544).


\end{document}